# Enhancement of Two-photon Absorption in Quantum Wells for Extremely Nondegenerate Photon Pairs

Himansu S. Pattanaik, Matthew Reichert, Jacob B. Khurgin, David J. Hagan, and Eric W. Van Stryland

**Abstract**

We recently demonstrated orders of magnitude enhancement of two-photon absorption (2PA) in direct gap semiconductors due to intermediate state resonance enhancement for photons of very different energies. It can be expected that further enhancement of nondegenerate 2PA will be observed in quantum wells (QW's) since the intraband matrix elements do not vanish near the band center as they do in the bulk, and the density of states in QW's is larger near the band edge. Here we present a perturbation-theory based theoretical description of nondegenerate 2PA in semiconductor QW's, where both frequency and polarization of two incident waves can vary independently. Analytical expressions for all possible permutations of frequencies and polarizations have been obtained, and the results are compared with degenerate 2PA in quantum wells along with degenerate and nondegenerate 2PA in bulk semiconductors. We show that using QW's in place of bulk semiconductors with both beams in the TM-polarized mode leads to an additional order of magnitude increase in the nondegenerate 2PA. Explicit calculations for GaAs QW's are presented.

*Index Terms*—**Nonlinear optics, Quantum wells.**

# I. INTRODUCTION

Two-photon absorption, 2PA, has a number of potential practical applications, such as photonic switching [1]–[5], optical limiting [6], [7], and others [8]–[11]. The study of 2PA is instrumental in investigating material properties, since two-photon transitions are subject to a set of selection rules that are distinct from those governing one-photon transitions [12]–[14]. Furthermore, nonlinear refraction, NLR, is related to nondegenerate 2PA though Kramers-Kronig relations, and NLR causes effects such as self-focusing [15], [16], self- and cross-phase modulation [17]–[20] and four-wave mixing [21], [22], each of them having a number of practical applications.

While the list of nonlinear optical materials investigated over the years is impressively long, semiconductors stand out among other materials since they have large nonlinearities, the technology is mature, high quality semiconductor materials are widely available, and the fabrication methods are well developed. Semiconductor nonlinear optical devices can be monolithically integrated with both passive and active photonic components as well as with electronic devices [23]–[25].

Two-photon absorption and NLR in semiconductors have been actively investigated by a number of groups providing theories yielding the scaling of the nonlinearities with bandgap [26]–[30]. Beginning in the 1980's, new epitaxial methods of growth led to the development of semiconductor structures with reduced dimensionality, i.e., quantum wells (QW's), which quickly showed their superiority to bulk semiconductors in such diverse applications as lasers, detectors, and modulators [8], [31]–[38]. The salient feature of QW's is the large density of states near the band edge, leading to the enhancement of both linear and nonlinear optical properties [5], [36]. These nonlinearities include 2PA, which has been studied extensively in the context of potential application in a variety of optical devices [39]–[41].

Most of the effort has been directed toward the study of degenerate two-photon absorption (D-2PA) properties of QW's where both photons are identical, i.e. they have the same energy and polarization. The applications of D-2PA include all-optical switching in the vicinity of the half-bandgap [42]–[45] where the nonlinear figure-of-merit is high, as well as detection and emission of entangled photon pairs [46]–[49].

The theoretical works by Spector et al [50] and Pasquarello and Quattropani [51] were among the first in predicting the spectral shape of D-2PA in QW's, but had discrepancies in the predicted spectra and the choice of gauge. The D-2PA theory by Shimizu *et al* considered excitonic effects and their predictions were confirmed by experimental observations [43], [52], [53]. Xia et al [54] discussed the D-2PA properties of QW's both in the presence and absence of external electric fields. Khurgin et al [42] developed analytical expressions for polarization dependent D-2PA in QW's., which was later expanded to include excitonic effects [55]. In this work we make modifications to the D-2PA theory presented in Ref. [42] and correct several calculations.

While D-2PA of QW's has been investigated both theoretically and experimentally, the nondegenerate two-photon absorption (ND-2PA), where two photons have different energies and polarizations, remains unexplored. In ND-2PA the energy of individual photons may approach intermediate-state resonances, thus leading to an enhancement of ND-2PA over D-2PA. In bulk semiconductors the large enhancement due to ND-2PA has been predicted theoretically [26], [30] and verified experimentally [56].

In the extremely nondegenerate (END) regime, when two photons are of extremely different (~10 ×) photon energies, an enhancement in the 2PA coefficient of two orders of magnitude has been measured in GaAs and ZnSe [56]. Based on this END enhancement, sensitive detection of mid-infrared (IR) femtosecond pulses in an uncooled p-i-n GaN photodiode has been demonstrated in Ref. [10]. The END-2PA in bulk semiconductors has also been used for detection of continuous wave IR laser light in uncooled GaAs p-i-n photodiodes [11], [57]. END enhancement is expected to be even more prominent in QW's near the band edge where the density of states and transition matrix elements exceed those of bulk semiconductors. For extreme nondegeneracy, the initial and final states must be near $k = 0$, where the density of states and the intraband transition matrix elements are small in the bulk but not necessarily small in quantum wells. It is also apparent that the END enhancement should exhibit a strong polarization dependence due to the anisotropy of the QW structure. In this work we develop a theoretical description of the spectrum and polarization dependence of ND-2PA in QW's. The greatest enhancement is found when using TM polarizations due to intermediate state resonance with intersubband transitions. Enhancement of one order of magnitude over bulk (three orders of magnitude over the degenerate case) should be experimentally feasible. We verify that in the limit of infinitely wide QW's that the 2PA becomes polarization independent and identical to that predicted in bulk semiconductors.

## II. THEORETICAL BACKGROUND

The enhancement in END-2PA in direct-gap semiconductors is similar to intermediate state resonance enhancement (ISRE) seen in molecules [12], [58], [59]. The ISRE occurs when the intermediate state for the transition approaches an eigenstate of the system and the 2PA is expected to diverge. The large enhancement for the END case over D-2PA can be understood qualitatively from the nondegenerate two-photon transition rate obtained from second-order perturbation theory,

$$W_2^{ND} = \frac{2\pi}{\hbar}\frac{2}{V}\sum_{vc}\left|\sum_i\left[\frac{\langle c|H_{2int}|i\rangle\langle i|H_{1int}|v\rangle}{E_{iv}(\mathbf{k}) - \hbar\omega_1} + \frac{\langle c|H_{1int}|i\rangle\langle i|H_{2int}|v\rangle}{E_{iv}(\mathbf{k}) - \hbar\omega_2}\right]\right|^2 \delta(E_{cv}(\mathbf{k}) - \hbar\omega_1 - \hbar\omega_2), \quad (1)$$

where $V$ is the volume, 1 and 2 represent indices for the two different photons, $v$ and $c$ represent valence band and conduction band respectively, $i$ represents the intermediate states, $\widehat{H}_{int}$ is the electron-field interaction Hamiltonian, $E_{cv}(\mathbf{k})$ is the energy difference between the valence and conduction bands, and $\hbar\omega_1$ and $\hbar\omega_2$ are the photon energies. This transition rate includes the spin degeneracy factor, is summed over all possible intermediate states $i$, and is also summed over all possible transitions starting from filled states in the valence band to empty states in the conduction band, i.e. a sum over electronic wave vector **k** and the bands. It has been predicted by Wherrett [27] and verified experimentally in Ref. [28] that allowed-forbidden transitions dominate 2PA in direct-gap semiconductors. The transition paths for the 2PA are shown in Fig. 1.

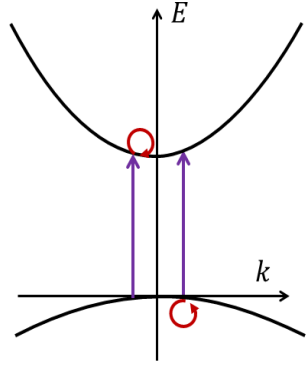

Fig. 1. Transition paths shown for ND-2PA in semiconductors corresponds to interband (line) and intraband (circle), or vice versa.

In END-2PA there are two resonances that can be exploited which lead to divergence of the expression in (1). The small energy photon can become near resonant to the "forbidden" or self-transition, i.e., zero energy resonance (intraband transition) while the large energy photon can be nearly resonant to the allowed one-photon (interband) transition. For semiconductors the intermediate states lie within the valence and conduction bands. For a two parabolic band model the intermediate state can either be in the valence band ($E_{iv}(\mathbf{k}) = 0$) or conduction band ($E_{iv}(\mathbf{k}) = \hbar\omega_1 + \hbar\omega_2$) and the nondegenerate transition rate can be written explicitly by considering all the possible transition paths (Fig. 1) for the system [56]

$$W_2^{ND} \propto \left| \frac{M_{cv}^2 M_{vv}^1}{-\hbar\omega_1} + \frac{M_{cv}^1 M_{vv}^2}{-\hbar\omega_2} + \frac{M_{cc}^1 M_{cv}^2}{\hbar\omega_1} + \frac{M_{cc}^2 M_{cv}^1}{\hbar\omega_2} \right|^2, \qquad (2)$$

where $M_{vv}^{1,2}$ and $M_{cc}^{1,2}$ are intraband and $M_{cv}^{1,2}$ are interband transition matrix elements respectively given by $M_{ji}^{1,2} = \langle j|H_{1,2_{int}}|i\rangle$. Note that the effective masses, and therefore the matrix elements, can have different signs.

An interesting fact observed looking at (1) is that each of the transition paths shown in Fig. 1 yield a term that diverges when one of the photon energies of the two-photon pair becomes small. The calculation of 2PA described here uses a two parabolic band model [26], [27]. A more detailed analysis for bulk 2PA using the Kane four band model has also been carried out [30]. Both of these models give similar results, which agree well with the experimentally measured values [56]. Thus for bulk semiconductors it is a very good approximation to use results of a two parabolic band model. Here we use a three-band model with parabolic bands, i.e. conduction band, heavy-hole band, and light-hole band for calculations of both D- and ND- 2PA in QW's. Since the calculation is carried out for undoped QW's, the transitions between the heavy-hole and light-hole bands are ignored [60]. Thus for 2PA coefficient calculations, we are really considering two separate two-band models coupled via the common conduction band.

The ND-2PA coefficient ($\alpha_2^{ND}$) can be derived by directly calculating the two-photon transition rate ($W_2^{ND}$) from second-order perturbation theory (1) by using the relation

$$\alpha_2^{ND}(\omega_1; \omega_2) = W_2^{ND} \frac{\hbar\omega_1}{2I_1 I_2}. \qquad (3)$$

Here $\alpha_2^{ND}(\omega_1; \omega_2)$ describes the absorption at $\omega_1$ induced by the presence of a field at $\omega_2$, and hence the irradiance change at $\omega_1$ is given by $dI_1/dz = -2\alpha_2^{ND}(\omega_1; \omega_2)I_2 I_1$ [56]. Note that $\alpha_2^{ND}(\omega_1; \omega_2)/\alpha_2^{ND}(\omega_2; \omega_1) = \omega_1/\omega_2$. As we will describe below, even greater enhancement of END-2PA due to ISRE is expected in semiconductor QW's.

### III. DEGENERATE TWO-PHOTON ABSORPTION IN QUANTUM WELLS

To calculate the D-2PA coefficient ($\alpha_2^D$), an ideal infinitely high barrier QW is considered. The width of the QW is $d$, with $x, y$ being the plane of the QW and $z$ the growth axis (Fig. 2). QW's have one dimensional confinement that lifts the degeneracy of the heavy-hole and light-hole bands at the band edge, and all bands split into many subbands. These subbands have confinement energies

$$E_{v,n} = \frac{n^2 \pi^2 \hbar^2}{2 m_{v,\perp} d^2}, \tag{4}$$

where $n = 1, 2, 3, ...$ represents the index for valence subbands or conduction subbands, $v = c, hh, lh$ represents conduction, heavy-hole, and light-holes respectively, $m_{v,\perp}$ is the longitudinal (perpendicular to the QW plane $x, y$) effective mass given by

$$m_{c,\perp} = m_c, \tag{5}$$

$$m_{hh,\perp} = \frac{m_0}{\gamma_1 - 2\gamma_2}, \tag{6}$$

and

$$m_{lh,\perp} = \frac{m_0}{\gamma_1 + 2\gamma_2}, \tag{7}$$

where $m_c$ and $m_0$ are the effective mass of the electron in the conduction band and the free electron mass respectively. $\gamma_1$ and $\gamma_2$ denote the Luttinger parameters [61]. Luttinger parameters are basically band structure parameters and are related to the curvature of the valence bands, hence related to effective mass in different **k** directions [61].

In the plane of the QW the dispersion is given by

$$E_{v,\parallel} = \frac{\hbar^2 \mathbf{k}_\parallel^2}{2 m_{v,\parallel}}, \tag{8}$$

where $\hbar \mathbf{k}_\parallel$ is the lateral quasi-momentum and $m_{v,\parallel}$ is the lateral (parallel to the QW plane $x, y$) effective mass given by

$$m_{c,\parallel} = m_c, \tag{9}$$

$$m_{hh,\parallel} = \frac{m_0}{\gamma_1 + \gamma_2}, \tag{10}$$

and

$$m_{lh,\parallel} = \frac{m_0}{\gamma_1 - \gamma_2}. \tag{11}$$

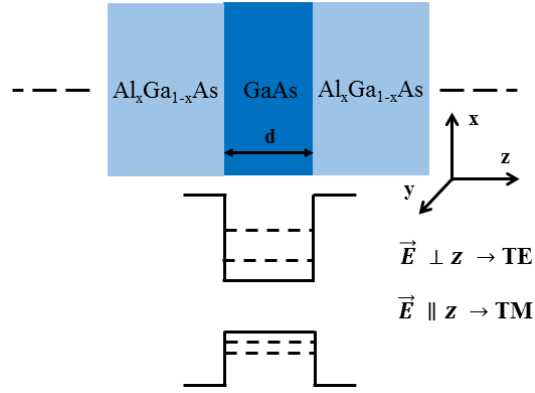

Fig. 2. Sketch of a finite QW structure showing electric vector polarization for TE and TM case.

Following the envelope approximation the wave function describing the sub-bands in a QW can be written as

$$|v,n\rangle = u_v \Psi_n(z) e^{i\mathbf{k}_\parallel \mathbf{r}_\parallel}, \qquad (12)$$

where $\mathbf{r}_\parallel = x\hat{\mathbf{x}} + y\hat{\mathbf{y}}$. For these infinitely high QW's the envelope wave function is given by

$$\Psi_n(z) = \sqrt{\frac{2}{d}} \sin\left(\frac{n\pi z}{d}\right), \qquad (13)$$

and the Bloch wave functions $u_v$ near the Brillouin zone center are given by

$$u_c = |iS \uparrow\rangle, \qquad (14)$$

$$u_{hh} = -\frac{1}{\sqrt{2}} |(X + iY) \uparrow\rangle, \qquad (15)$$

and

$$u_{lh} = -\frac{1}{\sqrt{6}} |(X + iY) \downarrow\rangle + \sqrt{\frac{2}{3}} |Z \uparrow\rangle, \qquad (16)$$

where $|S \uparrow\rangle$, is the band-edge function for the conduction band and $|X \uparrow\rangle, |Y \uparrow\rangle, |Z \uparrow\rangle$, are band edge functions for the valence band. The '↑' and '↓' represent the spin up and down of the electron or holes, respectively. We have similar band-edge functions for the down-spin of electrons or holes. These band-edge function are used to choose the basis functions, $|iS \uparrow\rangle, |-(X + iY)/\sqrt{2} \downarrow\rangle, |Z \uparrow\rangle, |(X - iY)/\sqrt{2} \downarrow\rangle$ and functions $|iS \downarrow\rangle, |(X - iY)/\sqrt{2} \uparrow\rangle, |Z \downarrow\rangle, |-(X + iY)/\sqrt{2} \uparrow\rangle$ used to form the heavy-hole, light-hole, and conduction band Bloch wave functions near the Brillouin zone center and calculate the corresponding eigen energies using the $\mathbf{k} \cdot \mathbf{p}$ method in Kane's model [62].

There are different approaches to the calculation of 2PA coefficients in a crystalline solid: 1) second-order perturbation theory [27], [29], [63] can be used to calculate the transition rate from valence band to conduction band. This theory predicts the D-2PA coefficients quite well using a simple two parabolic band model [26], [27]. The more complex four band model gives similar results [63]. 2) First-order perturbation theory can be used with a dressed-state electron wave function which includes acceleration of electrons as a result of the applied ac light field [64]. This method provides identical 2PA values and spectrum to those obtained from second-order perturbation theory [26], [29], [65]. 3) The 2PA coefficient can also be found from the imaginary part of the third-order susceptibility derived via perturbation theory [42]. In this paper we adopt the second-order perturbation theory, Fermi's Golden Rule approach to calculate the 2PA in an infinite QW. The two-photon transition rate per unit volume as derived using second-order perturbation theory is given by,

$$W_2^{ND} = \frac{2\pi}{\hbar}\frac{2}{V}\sum_{vc}\left|\sum_i \frac{\langle c|\hat{H}_{int}|i\rangle\langle i|\hat{H}_{int}|v\rangle}{E_{iv}(\mathbf{k}) - \hbar\omega}\right|^2 \times \delta(E_{cv}(\mathbf{k}) - 2\hbar\omega), \quad (17)$$

where $\hat{H}_{int} = -\frac{e}{m_0}\mathbf{A}(\mathbf{r},t)\cdot\hat{\mathbf{p}}$ represents the electron-field interaction Hamiltonian related to solids.

A QW is necessarily anisotropic, hence the 2PA rate depends on the direction of polarization of the incident light. For light polarized in the plane of the quantum well $\hat{e} \equiv \hat{x}$ or $\hat{y}$ (referred to as TE), the selection rules obey the condition $\Delta n = 0$ [13], [51], [52], [66]. For the TE-TE case, because $\Delta n = 0$, the transitions are necessarily of the allowed-forbidden type that occurs in bulk, and occur between valence and conduction bands of the same index $n$. Fig. 3 shows possible transition paths for 2PA with TE-TE polarized light.

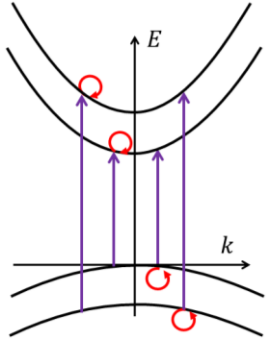

Fig. 3. Transitions shown here for two TE polarized waves corresponds to interband transition from valence subbands to conduction subbands and hole (or electron) intrasubband transitions.

Considering the selection rules for TE (i.e. TE-TE) polarized light, the transition rate is given by

$$W_2^D|_\parallel = \frac{2\pi}{\hbar}\frac{2}{V}\sum_v\sum_n\sum_{k_\parallel}\left|\frac{\langle c,n|\hat{H}_{int}|v,n\rangle\langle v,n|\hat{H}_{int}|v,n\rangle}{-\hbar\omega} + \frac{\langle c,n|\hat{H}_{int}|c,n\rangle\langle c,n|\hat{H}_{int}|v,n\rangle}{\hbar\omega}\right|^2 \quad (18)$$
$$\times \delta(E_{cv}(\mathbf{k}_\parallel) - 2\hbar\omega),$$

where the first summation is over 2PA contributions of light-holes and heavy-holes, the second is over subbands of the same index, and the third is over $\mathbf{k}_\parallel$ in the plane of the quantum well. As observed from the Bloch function (12), for TE polarized light there is a contribution to 2PA both from heavy-holes and light-holes, and the intrasubband matrix elements exist only for self transitions, i.e., transitions within a single subband, given by

$$\langle v|\mathbf{p}|v\rangle = \hbar\frac{m_0}{m_{v,\parallel}}\mathbf{k}_\parallel \text{ and}$$
$$\langle c|\mathbf{p}|c\rangle = \hbar\frac{m_0}{m_{c,\parallel}}\mathbf{k}_\parallel, \quad (19)$$

and the transition matrix elements for the intersubband matrix elements is given by

$$\langle c|\mathbf{p}|v\rangle = \mathbf{p}_{cv}, \quad (20)$$

Note from (19) that these matrix elements are proportional to $\mathbf{k}_\parallel$ showing how these transitions become weak (so-called "forbidden" transitions) as $|k_\parallel| \to 0$. Substituting the values of transition matrix elements from (19) and (20) and subband energies (4), transforming the summation over $\mathbf{k}_\parallel$ into an integration over kinetic energy, and using the momentum gauge for the interaction Hamiltonian, (18) is simplified to

$$W_2^D|_\parallel = \sum_v a_v |p_{cv}|^2 \frac{4}{\hbar^5} \frac{e^4 A_0^4}{16 m_0^4} \left(\frac{m_0}{\omega}\right)^2 \frac{1}{d} \sum_n (2\hbar\omega - E_{c,n} - E_{v,n} - E_g) \\ \times \Theta(2\hbar\omega - E_{c,n} - E_{v,n} - E_g), \tag{21}$$

where $\Theta(x)$ is the Heaviside step function, $a_v$'s are the coefficients obtained by averaging over the electron wave vector ($\mathbf{k}_\parallel$), and are equal to $1/4$ for heavy-holes and $1/12$ for light-holes. The D-2PA coefficient can be obtained from the transition rate by

$$\alpha_2^D|_\parallel = W_2^D|_\parallel \frac{2\hbar\omega}{I^2}, \tag{22}$$

where $I$ is the irradiance

$$I = \frac{n_\omega c \varepsilon_0 \omega^2 A_0^2}{2}, \tag{23}$$

The momentum matrix element $|p_{cv}|^2$ is related to the Kane energy parameter $E_p$ by

$$E_p = 2|p_{cv}|^2/m_0. \tag{24}$$

From (21) after the summation and using the relation (24), we obtain the following expression for the D-2PA coefficient for TE polarized light:

$$\alpha_2^D|_\parallel = 2\left(\frac{16\alpha}{n_\omega}\right)^2 \frac{2E_p (E_{lh}^{11})^2 \hbar d \mu_{lh,\perp}}{m_0 E_g^5} \left[\frac{1}{4}\frac{\mu_{lh,\perp}}{\mu_{hh,\perp}} F(\zeta_{hh}) + \frac{1}{12} F(\zeta_{lh})\right] \left(\frac{E_g}{2\hbar\omega}\right)^5, \tag{25}$$

where

$$\zeta_v = \frac{2\hbar\omega - E_g}{E_v^{11}}, \tag{26}$$

$$E_v^{11} = E_c^1 + E_v^1 = \frac{\hbar^2 \pi^2}{2\mu_{v,\perp} d^2}, \tag{27}$$

$$F(\zeta_v) = \left(\zeta_v N_v - \frac{1}{3} N_v^3 - \frac{1}{2} N_v^2 - \frac{1}{6} N_v\right), \tag{28}$$

$$N_v = Int(\sqrt{\zeta_v}), \tag{29}$$

and $\alpha = e^2/4\pi\varepsilon_0 \hbar c \approx 1/137$ is the fine structure constant, $n_\omega$ is the refractive index, $N_v$ is the number of two-photon transitions between the valence and conduction subbands. $\zeta_v$ defines how far above the bandgap ($E_g$) the two-photon transition energy is, normalized to the linear absorption edge.

For light polarized along the growth direction of the QW, $\hat{e} \equiv \hat{z}$ (referred as TM) the selection rule obeys the condition $\Delta n$ odd [13], [51], [52], [66]. For 2PA in the TM-TM case the interband transition involves a change of $n$ and the "intraband" transition occurs either between two conduction subbands or valence subbands. Fig. 4 shows some of the various transition paths possible for 2PA of TM-TM polarized light, which correspond to interband transitions from valence subbands to conduction subband and hole (or electron) intersubband transitions.

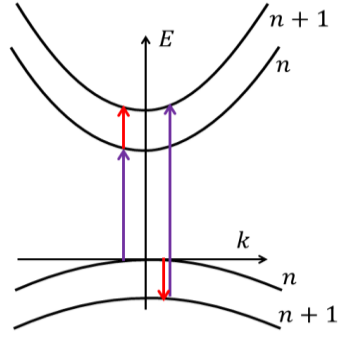

Fig. 4. Transitions shown here for two TM polarized waves correspond to interband transitions from the $nth$ valence subband to the $(n + 1)th$ conduction subband and hole (or electron) intersubband transitions.

Considering the selection rules for TM (i.e. TM-TM) polarized light, the transition rate is given by

$$W_2^D|_\perp = \frac{2\pi}{\hbar}\frac{2}{V}\sum_n \sum_{k_\parallel} \left| \frac{\langle c, n\pm 1|\hat{H}_{int}|c,n\rangle \langle c,n|\hat{H}_{int}|v,n\rangle}{\left[\hbar\omega - \frac{((n\pm 1)^2 - n^2)\pi^2\hbar^2}{2m_{c,\perp}d^2}\right]} \right.$$

$$\left. - \frac{\langle c, n\pm 1|\hat{H}_{int}|v,n\pm 1\rangle \langle v,n\pm 1|\hat{H}_{int}|v,n\rangle}{\left[\hbar\omega + \frac{((n\pm 1)^2 - n^2)\pi^2\hbar^2}{2m_{v,\perp}d^2}\right]} \right|^2 \delta(E_{cv}(\mathbf{k}_\parallel) - 2\hbar\omega). \quad (30)$$

The two-photon transitions in the TM-TM case are calculated in (32), only considering $\Delta n = \pm 1$. Higher odd $\Delta n$ values are ignored as their contribution is small [42]. Disregarding transitions between light-holes and heavy-holes [26] it can be determined from (12) that for TM-TM polarized light, contributions to 2PA occur only for light-hole and electron pairs. The intersubband matrix elements for TM-TM polarized light are given by

$$\langle v, n\pm 1|\mathbf{p}|v,n\rangle = -i\hbar \frac{m_0}{m_{v,\perp}} \frac{4(n\pm 1)n}{d((n\pm 1)^2 - n^2)} \hat{\mathbf{z}} \text{ and}$$

$$\langle c, n\pm 1|\mathbf{p}|c,n\rangle = -i\hbar \frac{m_0}{m_{c,\perp}} \frac{4(n\pm 1)n}{d((n\pm 1)^2 - n^2)} \hat{\mathbf{z}}. \quad (31)$$

Using the momentum gauge and following a similar method for simplification as carried out for the TE-TE case in (31), we obtain

$$W_2^D|_\perp = \frac{2\pi}{\hbar} \frac{e^4 A_0^4}{16 m_0^4} \frac{\mu_{lh,\parallel} m_0^2}{\pi d^3} \frac{2}{3} |p_{cv}|^2 \left[\frac{\hbar\omega}{\mu_{lh,\perp}}\right]^2 \left[\hbar\omega - \frac{(1\pm 2n)\pi^2\hbar^2}{2m_{c,\perp}d^2}\right]^{-2}$$

$$\times \left[\hbar\omega + \frac{(1\pm 2n)\pi^2\hbar^2}{2m_{lh,\perp}d^2}\right]^{-2} \Theta(2\hbar\omega - E_g - E_{c,n\pm 1} - E_{lh,n}). \quad (32)$$

After simplification of (32) and using the relation in (24) we obtain the following expression for the D-2PA coefficient for TM-TM polarized light:

$$\alpha_2^D|_\perp = \frac{1}{3}\left(\frac{32\alpha}{\pi n_\omega}\right)^2 \frac{2E_p(E_{lh}^{11})^2 d\mu_{lh,\parallel}\hbar}{m_0 E_g^5} \left(\frac{E_g}{2\hbar\omega}\right)^5 [F(N_1) + F(N_2)], \quad (33)$$

where

$$F(N_{1(2)}) = \sum_n \left[ \left[ (2n+1) - \frac{1}{(2n+1)} \right]^2 \left[ \hbar\omega - \frac{(1+2n)\pi^2\hbar^2}{2m_{c(lh,\perp)}d^2} \right]^{-2} \left[ \hbar\omega + \frac{(1+2n)\pi^2\hbar^2}{2m_{lh,\perp(c)}d^2} \right]^{-2} \right], \quad (34)$$

and

$$N_{1(2)} = Int\left[ \left( \zeta_{lh} - \frac{\mu_{lh,\perp}}{m_{c(lh,\perp)}} + \left( \frac{\mu_{lh,\perp}}{m_{c(lh,\perp)}} \right)^2 \right)^{1/2} - \frac{\mu_{lh,\perp}}{m_{c(lh,\perp)}} \right], \quad (35)$$

where $N_1$ is the number of all possible two-photon transitions from the $nth$ light-hole subband to the $(n+1)th$ conduction subband, and there are $N_2$ possible two-photon transitions from the $(n+1)th$ light-hole subband to the $nth$ conduction subband.

Fig. 5 shows results for the D-2PA coefficient plotted against the two-photon energy ($2\hbar\omega$) for the TE-TE and TM-TM case calculated for an infinitely high barrier in a GaAs QW of thickness 10 nm, along with the D-2PA coefficient for bulk GaAs calculated using the relation [26], [27]

$$\alpha_2^D = \left( \frac{16\alpha}{n_\omega} \right)^2 \frac{1}{5} \frac{\pi E_p \hbar^2}{m_0} \frac{2^{3/2}\mu_r^{1/2}}{E_g^{7/2}} F(x), \quad (36)$$

where

$$F(x) = \frac{(2x-1)^{3/2}}{(2x)^5} \quad \text{and} \quad x = \frac{\hbar\omega}{E_g}, \quad (37)$$

$E_p$ is the Kane energy and $\mu_r = (1/m_c + 1/m_{hh})^{-1}$ is the reduced mass of electron and heavy-hole. Note that $E_p$ scales as $E_g^{-1/2}$ so that $\alpha_2^D$ scales as $E_g^{-3}$. The material parameters for GaAs used in the calculations to generate Fig. 5, and all subsequent figures in this paper, are given in Table I.

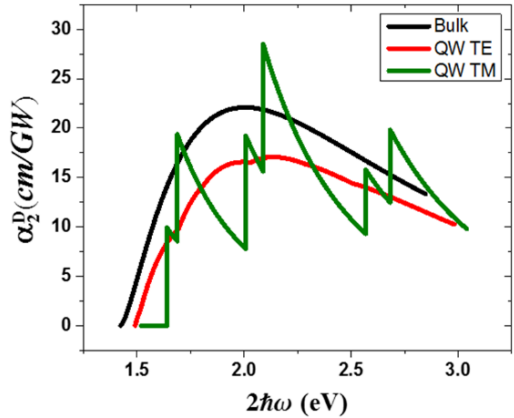

Fig. 5. D-2PA coefficient ($\alpha_2^D$) for bulk GaAs and GaAs QW (TE-TE and TM-TM) of width 10 $nm$.

The D-2PA coefficient $\alpha_2^D|_\parallel$ has discontinuities in the derivative of the D-2PA curve as we cross the transition from the $nth$ valence subband to the $nth$ conduction subband. The continuous increase of $\alpha_2^D|_\parallel$ from one band to the other band is due to the linear dependence of the intraband transition matrix elements on the in-plane wave vector (**k**) (19). We do not observe a step-like increase in $\alpha_2^D|_\parallel$ as observed in one-photon absorption in QW's, because when the two-photon energy increases just past the $nth$ transition energy, the in-plane wave vector is zero. In the TM-TM case the D-2PA coefficient $\alpha_2^D|_\perp$ is more structured and shows the step-like density of states features of a QW corresponding to each band-to-band transition. $\alpha_2^D|_\perp$ shows step-like features because the intraband matrix elements are **k** independent (31).

TABLE I
MATERIAL PARAMETERS FOR GaAs [74], [75]

| Parameters | Value |
|---|---|
| $E_g$ (eV) | 1.424 |
| $E_p$ (eV) | 28.8 |
| $m_c/m_0$ | 0.067 |
| $m_{hh}/m_0$ | 0.5 |
| $\gamma_1$ | 6.8 |
| $\gamma_2$ | 1.9 |

Fig. 6 shows the D-2PA coefficient evaluated for different thicknesses for the TE-TE and TM-TM cases along with that for bulk GaAs. In these plots, the sum of the photon energies is shown scaled to the respective linear absorption edge energies of the bulk and QW since this allows comparing bulk and QW, semiconductors on the same scale. For the bulk case, $\alpha_2^D$ is plotted against the two-photon energy normalized to the bandgap, whereas for the QW, $\alpha_2^D|_\parallel$ is plotted against the two-photon energy normalized to $E_g + E_{hh}^{11}$ for the TE-TE case, and $\alpha_2^D|_\perp$ is plotted against the two-photon energy normalized to $E_g + E_{lh}^{11}$ for the TM-TM case.

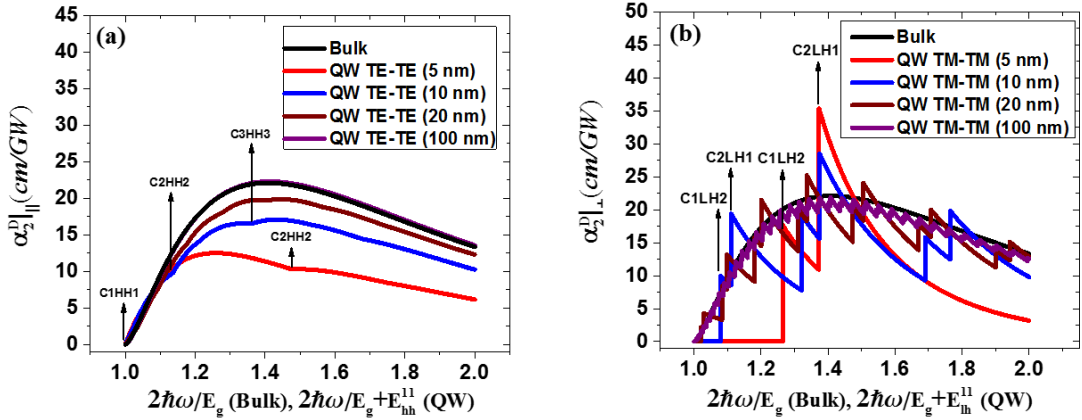

Fig. 6. D-2PA coefficient in bulk GaAs and GaAs QW's of different widths: (a) TE-TE case and (b) TM-TM case. The arrows indicate valence band to conduction band transition energies.

As we go higher in two-photon energy ($2\hbar\omega$), $\alpha_2^D$ increases for photon energies close to the middle of the bandgap, goes through a maximum and then decreases for photon energies close to the bandgap. The initial increase in $\alpha_2^D$ for midgap photon energies is attributed to the dominant contribution of the increased density of states, and the decrease for photon energies close to the bandgap is attributed to the dominant contribution of the increase in the detuning term $E_{iv}(\mathbf{k}) - \hbar\omega$ (see (17)). For the QW, we observe that $\alpha_2^D|_\parallel$ decreases with a decrease in the QW width ($d$). This can be explained by the increase in the effective gap energy due to confinement. For the QW in the TM-TM case we observe regions with an increase in $\alpha_2^D|_\perp$ with a decrease in $d$. This modulation of the 2PA curve is due to the step-like features of the density of states. The arrows in Fig. 6 label the valence subband to conduction subband transition energies. As discussed earlier, for the TE-TE case only $\Delta n = 0$, and for the TM-TM case $\Delta n =$ $\pm 1$, are considered. In the TM-TM case, the first transition is from the 2$^{nd}$ light-hole subband to the 1$^{st}$ electron subband, thus the onset of 2PA occurs at $2\hbar\omega = E_g + E_{lh}^{12}$, rather than $2\hbar\omega = E_g + E_{lh}^{11}$. In the limiting case of a wide QW, both $\alpha_2^D|_\parallel$ and $\alpha_2^D|_\perp$ approach the bulk value $\alpha_2^D$ as seen in Fig. 6.

## IV. NONDEGENERATE TWO-PHOTON ABSORPTION IN QUANTUM WELLS

For ND-2PA, the transition rate obtained from second-order perturbation theory, where each photon of energies $\hbar\omega_1$ and $\hbar\omega_2$ is absorbed, is given by,

$$W_2^{ND} = \frac{2\pi}{\hbar}\frac{2}{V}\sum_{vc}\left|\sum_i\left[\frac{\langle c|H_{2_{int}}|i\rangle\langle i|H_{1_{int}}|v\rangle}{E_{iv}(\mathbf{k})-\hbar\omega_1}+\frac{\langle c|H_{1_{int}}|i\rangle\langle i|H_{2_{int}}|v\rangle}{E_{iv}(\mathbf{k})-\hbar\omega_2}\right]\right|^2 \tag{38}$$
$$\times \delta(E_{cv}(\mathbf{k})-\hbar\omega_1-\hbar\omega_2),$$

where $\widehat{H}_{1\,int} = -\frac{e}{m_0}\mathbf{A}_1(\mathbf{r},t)\cdot\widehat{\mathbf{p}}$ and $\widehat{H}_{2\,int} = -\frac{e}{m_0}\mathbf{A}_2(\mathbf{r},t)\cdot\widehat{\mathbf{p}}$ represent the electron-field interaction Hamiltonian related to solids. Similar to the degenerate case, considering the selection rules for TE (i.e. TE-TE) polarized light, the transition rate for the ND-2PA for the TE-TE case may be simplified to yield

$$W_2^{ND}|_\parallel = \frac{2\pi}{\hbar}\frac{2}{V}\sum_v\sum_n\sum_{k_\parallel}\left|\frac{\langle c,n|\widehat{H}_{2\,int}|v,n\rangle\langle v,n|\widehat{H}_{1\,int}|v,n\rangle}{-\hbar\omega_1}\right.$$
$$+\frac{\langle c,n|\widehat{H}_{1\,int}|v,n\rangle\langle v,n|\widehat{H}_{2\,int}|v,n\rangle}{-\hbar\omega_2}+\frac{\langle c,n|\widehat{H}_{1\,int}|c,n\rangle\langle c,n|\widehat{H}_{2\,int}|v,n\rangle}{\hbar\omega_1} \tag{39}$$
$$\left.+\frac{\langle c,n|\widehat{H}_{2\,int}|c,n\rangle\langle c,n|\widehat{H}_{1\,int}|v,n\rangle}{\hbar\omega_2}\right|^2\delta(E_{cv}(\mathbf{k}_\parallel)-\hbar\omega_1-\hbar\omega_2).$$

The summations over $v$, $n$, and $k$ are the same as described for (18). As mentioned in the calculations of D-2PA for the TE-TE case, there are contributions both from heavy-holes and light-holes to 2PA for TE polarized light and the intersubband matrix elements exist only for transitions within a single subband. Substituting values for the transition matrix elements (19) and subband energies (4) and transforming the summation over the in-plane wave vector $\mathbf{k}_\parallel$ into an integration over kinetic energy, we obtain

$$W_2^{ND}|_\parallel = \sum_v a_v|p_{cv}|^2\frac{4}{\hbar^5}\frac{e^4 A_{01}^2 A_{02}^2}{16 m_0^2}\frac{1}{d}\left(\frac{1}{\omega_1}+\frac{1}{\omega_2}\right)^2\sum_n(\hbar\omega_1+\hbar\omega_2-E_{e,n}-E_{v,n}-E_g) \tag{40}$$
$$\times \Theta(\hbar\omega_1+\hbar\omega_2-E_{e,n}-E_{v,n}-E_g),$$

After the summations in (40) and substituting $E_p$ from (24) into (40), we obtain the following expression for the ND-2PA coefficient for the TE-TE polarized light:

$$\alpha_2^{ND}(\omega_1;\omega_2)|_\parallel = 2\frac{(16\alpha)^2}{n_{\omega_1}n_{\omega_2}}\frac{2E_p(E_{lh,11})^2 d\mu_{lh,\perp}}{m_0 E_g^5}\left[\frac{1}{4}\frac{\mu_{lh,\perp}}{\mu_{hh,\perp}}F(\zeta_{hh}^{ND})+\frac{1}{12}F(\zeta_{lh}^{ND})\right]$$
$$\times\frac{\hbar}{2^7\frac{\hbar\omega_1}{E_g}\left(\frac{\hbar\omega_2}{E_g}\right)^2}\left(\frac{E_g}{\hbar\omega_1}+\frac{E_g}{\hbar\omega_2}\right)^2, \tag{41}$$

where for the nondegenerate case

$$\zeta_v^{ND} = \frac{\hbar\omega_1+\hbar\omega_2-E_g}{E_v^{11}}. \tag{42}$$

The expressions for $E_v^{11}$, $F(\zeta_v^{ND})$, and $N_v$ in the nondegenerate case have the same form as in (27), (28), and (29) except that $\zeta_v^{ND}$ is now given by (42). Here $N_v$ is the number of nondegenerate two-photon transitions between the valence and conduction subbands.

Considering the selection rules, the transition rate for ND-2PA for TM (i.e. TM-TM) polarized light, is given by

$$W_2^{ND}|_\perp = \frac{2\pi}{\hbar} \frac{2}{V} \sum_n \sum_{k_\parallel} \left| \frac{\langle c, n \pm 1|\hat{H}_{2_{int}}|v, n \pm 1\rangle \langle v, n \pm 1|\hat{H}_{1_{int}}|v, n\rangle}{-\left[\hbar\omega_1 + \frac{((n \pm 1)^2 - n^2)\pi^2\hbar^2}{2m_{lh,\perp}d^2}\right]} \right.$$

$$+ \frac{\langle c, n \pm 1|\hat{H}_{1_{int}}|v, n \pm 1\rangle \langle v, n \pm 1|\hat{H}_{2_{int}}|v, n\rangle}{-\left[\hbar\omega_2 + \frac{((n \pm 1)^2 - n^2)\pi^2\hbar^2}{2m_{lh,\perp}d^2}\right]} + \frac{\langle c, n \pm 1|\hat{H}_{2_{int}}|c, n\rangle \langle c, n|\hat{H}_{1_{int}}|v, n\rangle}{\hbar\omega_2 - \frac{((n \pm 1)^2 - n^2)\pi^2\hbar^2}{2m_{c,\perp}d^2}} \quad (43)$$

$$\left. + \frac{\langle c, n \pm 1|\hat{H}_{1_{int}}|c, n\rangle \langle c, n|\hat{H}_{2_{int}}|v, n\rangle}{\hbar\omega_1 - \frac{((n \pm 1)^2 - n^2)\pi^2\hbar^2}{2m_{c,\perp}d^2}} \right|^2 \times \delta(E_{cv}(\mathbf{k}_\parallel) - \hbar\omega_1 - \hbar\omega_2).$$

Substituting intersubband and interband transition matrix elements from (20) and (31), (43) becomes

$$W_2^{ND}|_\perp = \frac{2\pi}{\hbar} \frac{e^4 A_{01}^2 A_{02}^2}{16 m_0^4} \frac{\hbar^2 m_0^2}{(\mu_{lh,\perp})^2 d^2} \sum_n \frac{2}{V} \sum_{k_\parallel} \left[\frac{4(n \pm 1)n}{1 \pm 2n}\right]^2 \frac{2}{3} |p_{cv}|^2$$

$$\times \left[ \frac{\hbar\omega_1}{\left[\hbar\omega_1 - \frac{(1 \pm 2n)\pi^2\hbar^2}{2m_{c,\perp}d^2}\right]\left[\hbar\omega_1 + \frac{(1 \pm 2n)\pi^2\hbar^2}{2m_{lh,\perp}d^2}\right]} \right. \quad (44)$$

$$\left. + \frac{\hbar\omega_2}{\left[\hbar\omega_2 - \frac{(1 \pm 2n)\pi^2\hbar^2}{2m_{c,\perp}d^2}\right]\left[\hbar\omega_2 + \frac{(1 \pm 2n)\pi^2\hbar^2}{2m_{lh,\perp}d^2}\right]}\right]^2 \times \delta(E_{cv}(\mathbf{k}_\parallel) - \hbar\omega_1 - \hbar\omega_2).$$

The summation over $n$ corresponds to allowed two-photon transitions between valence subbands $n$ and conduction subbands $n \pm 1$. Simplifying (44) we obtain

$$W_2^{ND}|_\perp = \frac{2\pi}{\hbar} \frac{e^4 A_{01}^2 A_{02}^2}{16 m_0^4} \frac{\hbar^2 m_0^2}{(\mu_{lh,\perp})^2 d^2} \frac{\mu_{lh,\parallel}}{\pi \hbar^2 d} \frac{2}{3} |p_{cv}|^2 \sum_n [F(N_1) + F(N_2)], \quad (45)$$

where $F(N_1)$ and $F(N_2)$ are given by

$$F(N_{1(2)}) = \left| \left[(2n+1) - \frac{1}{(2n+1)}\right] \times \left[ \frac{\hbar\omega_1}{\left[\hbar\omega_1 - \frac{(1+2n)\pi^2\hbar^2}{2m_{c(lh,\perp)}d^2}\right]\left[\hbar\omega_1 + \frac{(1+2n)\pi^2\hbar^2}{2m_{lh,\perp(c)}d^2}\right]} \right. \right.$$

$$\left. \left. + \frac{\hbar\omega_2}{\left[\hbar\omega_2 - \frac{(1+2n)\pi^2\hbar^2}{2m_{c(lh,\perp)}d^2}\right]\left[\hbar\omega_2 + \frac{(1+2n)\pi^2\hbar^2}{2m_{lh,\perp(c)}d^2}\right]}\right]\right|^2. \quad (46)$$

The ND-2PA coefficient for TM-TM polarized light is obtained from the transition rate as

$$\alpha_2^{ND}(\omega_1;\omega_2)|_\perp = \frac{1}{3}\frac{\left(\frac{32\alpha}{\pi}\right)^2}{n_{\omega_1}n_{\omega_2}}\frac{2E_p(E_{lh,11})^2\mu_{lh,\parallel}\hbar d}{m_0 E_g^3} \times \sum_n [F(N_1)+F(N_2)]\frac{1}{2^7\frac{\hbar\omega_1}{E_g}\left(\frac{\hbar\omega_2}{E_g}\right)^2}. \quad (47)$$

Fig. 7 shows results for the ND-2PA coefficient of bulk GaAs and for a GaAs QW having infinitely high barriers and of different widths for both the TE-TE and TM-TM case calculated at optical frequencies $\omega_1$ and $\omega_2$. The ND-2PA coefficient of bulk GaAs is calculated using the relation [26], [30]

$$\alpha_2^{ND}(\omega_1;\omega_2) = \frac{(16\alpha)^2}{n_{\omega_1}n_{\omega_2}}\frac{1}{5}\frac{\pi E_p \hbar^2 2^{3/2}\mu_r^{1/2}}{m_0 E_g^{7/2}} F(x_1,x_2), \quad (48)$$

where

$$F(x_1;x_2) = \frac{(x_1+x_2-1)^{3/2}}{2^7 x_1 x_2^2}\left(\frac{1}{x_1}+\frac{1}{x_2}\right)^2,$$
$$x_1 = \frac{\hbar\omega_1}{E_g}, x_2 = \frac{\hbar\omega_2}{E_g}. \quad (49)$$

In Fig. 7 the ND-2PA coefficients are plotted against the photon energies scaled with respect to the one-photon transition energies. This allows comparison of the ND-2PA coefficient for the bulk and QW semiconductors on the same scale and also makes comparison to the respective D-2PA coefficient easier. Due to the large energy difference of photon pairs in the nondegenerate case, for a bulk semiconductor there is about a hundred-fold increase in $\alpha_2^{ND}(\omega_1;\omega_2)$ over $\alpha_2^D$. This enhancement has been measured for direct-gap semiconductors [56] and has been used for different applications such as mid-infrared (mid-IR) detection [10] and imaging [67].

In a standard experimental setup, the measurement of $\alpha_2^{ND}(\omega_1;\omega_2)$, such as that carried out in Ref. [56], uses a pump-probe geometry where the transmission of the weak probe beam at $\omega_1$ is monitored in the presence of the strong pump beam at $\omega_2$. The pump is chosen to have a long wavelength to avoid any D-2PA or three-photon absorption of the pump itself. The enhancement in $\alpha_2^{ND}(\omega_1;\omega_2)$ is explained by the resonant terms in the denominator of (38).

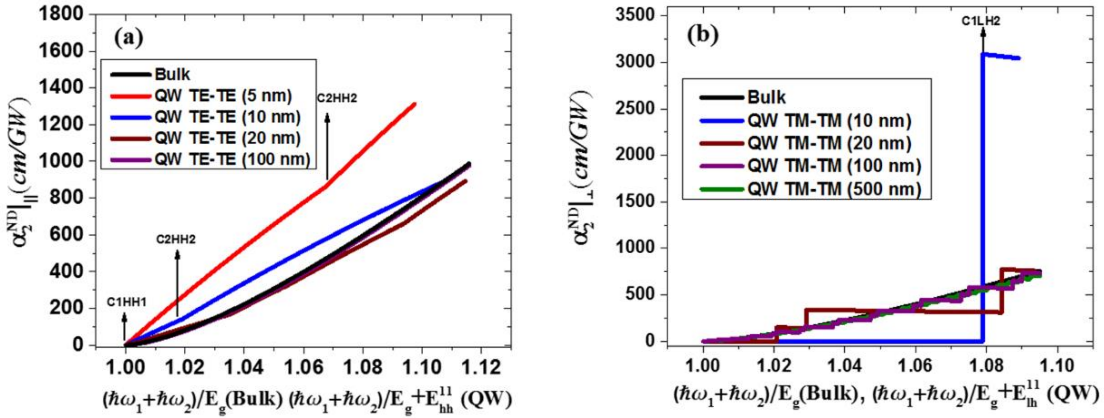

Fig. 7. ND-2PA coefficient in bulk GaAs and GaAs QW's of different widths: (a) TE-TE case and (b) TM-TM case. The arrows indicate valence band to conduction band transition energies.

Both plots in Fig. 7 are generated with a pump photon energy $\hbar\omega_2 \approx 0.12 E_g$, corresponding to a wavelength of 7.5 μm and varying the probe photon energy $\hbar\omega_1$. In Fig. 7, and all subsequent plots of 2PA versus energy, we restrict the probe photon energy, $\hbar\omega_1$, to be 30 meV below the linear absorption edge. This is done to keep the linear Urbach–tail absorption low. For the TM-TM case the linear absorption edge occurs at $E_g + E_{lh}^{11}$. For the QW in the TE-TE case, $\alpha_2^{ND}|_\parallel$ shows similar continuous transitions from the $nth$ valence subband to the $nth$ conduction subband as in D-2PA. The transition paths for ND-2PA in the TE-TE and TM-TM cases are shown in Fig. 3 and Fig. 4 respectively.

Fig. 7 (a) shows $\alpha_2^{ND}|_\parallel(\omega_1;\omega_2)$ for QW widths of $10\ nm$ and $5\ nm$. We observe a small increase in $\alpha_2^{ND}|_\parallel(\omega_1;\omega_2)$ over the bulk for larger confinements, attributed to the dominant contribution of the increased density of states in QW's. For $(\hbar\omega_1 + \hbar\omega_2)/E_g = (\hbar\omega_1 + \hbar\omega_2)/(E_g + E_{hh}^{11}) = 1.02$, $\alpha_2^{ND}|_\parallel(\omega_1;\omega_2)$ for a QW of width $10\ nm$ is $\approx 2$ times the bulk $\alpha_2^{ND}(\omega_1;\omega_2)$ and for a QW of width $5\ nm$ is $\approx 3.4$ times the bulk $\alpha_2^{ND}(\omega_1;\omega_2)$. Similar to $\alpha_2^{D}|_\parallel$ there is a continuous increase of $\alpha_2^{ND}|_\parallel(\omega_1;\omega_2)$ due to the linear dependence of the intraband transition matrix elements on the in-plane wave vector ($\mathbf{k}_\parallel$) (19) and the signature step-like feature seen in the linear absorption spectrum of QW's is not observed.

For ND-2PA in the TM-TM case (Fig. 7 (b)) $\alpha_2^{ND}|_\perp(\omega_1;\omega_2)$ shows more structured features, as in D-2PA, due to the similar selection rules and transition paths. For $(\hbar\omega_1 + \hbar\omega_2)/E_g = 1.02$ in the bulk case, $\alpha_2^{ND}(\omega_1;\omega_2) \approx 82\ cm/GW$ and for $(\hbar\omega_1 + \hbar\omega_2)/(E_g + E_{lh}^{12}) = 1.02$, in a QW of width $10\ nm$, $\alpha_2^{ND}|_\perp(\omega_1;\omega_2) \approx 3000\ cm/GW$ which is $\approx 36\times$ that of the bulk.

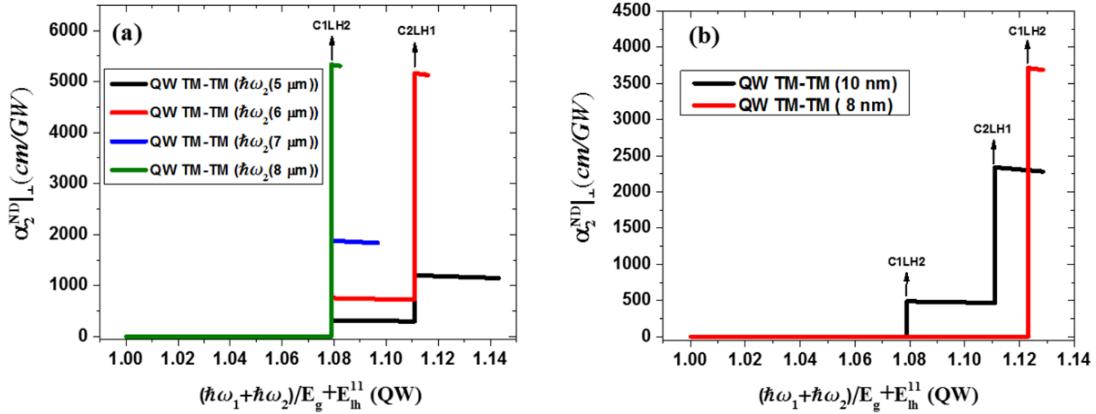

Fig. 8. (a) ND-2PA coefficient ($\alpha_2^{ND}|_\perp$) in a GaAs QW of width $10\ nm$ for various pump photon energies. (b) ND-2PA coefficient ($\alpha_2^{ND}|_\perp$) in a GaAs QW of different widths for a constant pump photon energy $\hbar\omega_2$ corresponding to a wavelength of 5.5 μm. The arrows indicate valence band to conduction band transition energies.

Since in the TM-TM case, the 1st two-photon transition occurs at $(E_g + E_{lh}^{12})$, the range over which the pump ($\hbar\omega_2$) and probe ($\hbar\omega_1$) photon energies can vary depends on the quantum well width and gives a limit for the extreme nondegenerate cases for QW's with large confinement. The photon energy condition for the pump is given by,

$$\hbar\omega_2 > E_{lh}^2 - E_{lh}^1. \tag{50}$$

The range of probe photon energies is given by,

$$E_g + E_{lh}^{12} - \hbar\omega_2 < \hbar\omega_1 < E_g + E_{lh}^{11}. \tag{51}$$

This is depicted in Fig. 8, which shows different scenarios for $\alpha_2^{ND}|_\perp(\omega_1;\omega_2)$. In Fig. 8 (a), $\alpha_2^{ND}|_\perp(\omega_1;\omega_2)$ is plotted for a QW of width $10\ nm$ for different values of $\hbar\omega_2$, while $\hbar\omega_1$ is varied. In Fig. 8 (b) $\alpha_2^{ND}|_\perp(\omega_1;\omega_2)$ is plotted for QW's of different widths with a constant $\hbar\omega_2$ while varying $\hbar\omega_1$.

In Fig. 8 (a) as we decrease the photon energy $\hbar\omega_2$, we observe a strong increase in $\alpha_2^{ND}|_\perp(\omega_1;\omega_2)$ at the C1LH2 transition. The C2LH1 transitions are not observed for $\hbar\omega_2(8\ \mu m)$ and $\hbar\omega_2(7\ \mu m)$, because $\hbar\omega_1 + \hbar\omega_2 < E_g + E_{lh}^{12}$. This increase is also observed in Fig. 8 (b) as we decrease the QW width from $10\ nm$ to $8\ nm$. As the confinement increases there is a strong enhancement of $\alpha_2^{ND}|_\perp(\omega_1;\omega_2)$ at the C1LH2 transition, but the C2LH1 transition is not observed (Fig. 8 (b)) for $\hbar\omega_2(5.5\ \mu m)$.

The above cases are examined for ND-2PA in a QW in which the two-photons are of the same polarization. It is also interesting to investigate the case of ND-2PA where one of the incident beams is TE polarized and the other beam is TM polarized. We refer this as the TE-TM case. We let the TE polarized wave have photon energy $\hbar\omega_1$ while the TM polarized wave has photon energy $\hbar\omega_2$. The transition paths for ND-2PA in the TE-TM case are shown in Fig. 9.

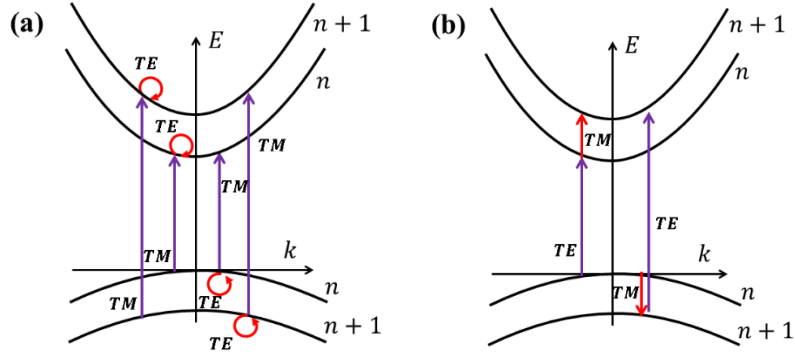

Fig. 9. (a) Transitions shown here corresponds to an interband transition from a valence subband to conduction subband and a light-hole (or electron) intrasubband transition, and (b) Transitions shown here correspond to an interband transition from the $nth$ valence subband to the $(n+1)th$ conduction subband and a light-hole (or electron) or heavy-hole (or electron) intersubband transition.

In Fig. 9 (a) the transition paths are shown only for light-hole contributions to the 2PA as the heavy-hole interband transition is not excited by TM polarized light. In Fig. 9 (b), the transition paths shown have both light-hole and heavy-hole contributions to the 2PA. This is because TE polarized light excites both heavy-hole and light-hole interband transitions and TM polarized light excites both heavy-hole and light-hole intersubband transitions. Considering the two-photon transition paths shown in Fig. 9 (following the appropriate selection rules) the two-photon transition rate for ND-2PA for the TE-TM case is given by

$$W_2^{ND}|_{TE-TM} = \frac{2\pi}{\hbar}\frac{e^4 A_{01}^2 A_{02}^2}{16 m_0^4}\sum_n \frac{2}{V}\sum_{k_\parallel} \qquad (52)$$

$$\left| \frac{\hat{m}_2 \cdot \mathbf{p}_{cv} \hat{e}_1 \cdot \hbar \frac{m_0}{m_{v,\|}} \mathbf{k}_\|}{-\hbar\omega_1} + \frac{\hat{e}_1 \cdot \hbar \frac{m_0}{m_e} \mathbf{k}_\| \hat{m}_2 \cdot \mathbf{p}_{cv}}{\hbar\omega_1} \right|^2 \delta(E_{cv}(\mathbf{k}_\|) - \hbar\omega_1 - \hbar\omega_2)$$

$$+ \frac{2\pi}{\hbar} \frac{e^4 A_{01}^2 A_{02}^2}{16 m_0^4} \sum_v \sum_n \frac{2}{V} \sum_{k_\|}$$

$$\left| \frac{\hat{e}_1 \cdot \mathbf{p}_{cv} - i\hbar \frac{m_0}{m_{v,\perp}} \frac{4(n\pm 1)n}{d(1\pm 2n)} \hat{m}_2 \cdot \hat{\mathbf{z}}}{-\left[\hbar\omega_2 + \frac{\left(n^{\,2} - n^2\right)\pi^2 \hbar^2}{2m_{v,\perp}d^2}\right]} + \frac{-i\hbar \frac{m_0}{m_e} \frac{4(n\pm 1)n}{d(1\pm 2n)} \hat{m}_2 \cdot \hat{\mathbf{z}} \hat{e}_1 \cdot \mathbf{p}_{cv}}{\hbar\omega_2 - \frac{\left(n^{\,2} - n^2\right)\pi^2 \hbar^2}{2m_{e,\perp}d^2}} \right|^2$$

$$\times \delta(E_{cv}(\mathbf{k}_\|) - \hbar\omega_1 - \hbar\omega_2),$$

where $\hat{e}_1$ and $\hat{m}_2$ represents polarization of TE and TM beams respectively, and summation over $v$ corresponds to contributions of light-holes and heavy-holes to ND-2PA. Following the procedure carried out in the derivation of the ND-2PA coefficient of TE and TM cases, $\alpha_2^{ND}|_{TE-TM}(\omega_1; \omega_2)$ is given by,

$$\alpha_2^{ND}|_{TE-TM}(\omega_1; \omega_2) = \frac{\alpha^2}{n_{\omega_1} n_{\omega_2}} \frac{E_p}{m_0} \frac{8\hbar d}{\pi^2 E_g^3}$$

$$\times \left[ \left[\frac{E_g}{\hbar\omega_1}\right]^2 (E_{lh}^{11})^2 \frac{\mu_{lh,\perp} \pi^2}{E_g^2} \frac{1}{3} F(\zeta_{lh}) \right.$$

$$+ (E_{lh}^{11})^2 \frac{4\hbar^2 d}{\pi^2} \left[ \frac{1}{6} \mu_{lh,\|} \sum_n [F_{lh}(N_1) + F_{lh}(N_2)] \right.$$

$$\left. \left. + \frac{(\mu_{lh,\perp})^2}{(\mu_{hh,\perp})^2} \frac{1}{2} \mu_{hh,\|} \sum_n [F_{hh}(N_1) + F_{hh}(N_2)] \right] \right] \frac{E_g^3}{(\hbar\omega_1)(\hbar\omega_2)^2}, \quad (53)$$

where $N_1$ and $N_2$ are given by (35) and $F_{lh}(N_{1(2)})$ and $F_{hh}(N_{1(2)})$ are given by

$$F_{lh}(N_{1(2)}) = \left[ \left[(2n+1) - \frac{1}{(2n+1)}\right] \right.$$

$$\left. \times \hbar\omega_2 \left[\hbar\omega_2 - \frac{(1+2n)\pi^2 \hbar^2}{2m_{c(lh,\perp)}d^2}\right]^{-1} \left[\hbar\omega_2 + \frac{(1+2n)\pi^2 \hbar^2}{2m_{lh,\perp(c)}d^2}\right]^{-1} \right]^2, \quad (54)$$

and

$$F_{hh}(N_{1(2)}) = \left[ \left[(2n+1) - \frac{1}{(2n+1)}\right] \right.$$

$$\left. \times \hbar\omega_2 \left[\hbar\omega_2 - \frac{(1+2n)\pi^2 \hbar^2}{2m_{c(hh,\perp)}d^2}\right]^{-1} \left[\hbar\omega_2 + \frac{(1+2n)\pi^2 \hbar^2}{2m_{hh,\perp(c)}d^2}\right]^{-1} \right]^2. \quad (55)$$

In Fig. 10 the ND-2PA coefficients evaluated for bulk GaAs and GaAs QW's for the TM-TM case and TE-TM case are shown for a pump photon energy $\hbar\omega_2(7.5\ \mu m)$. For the QW, $\alpha_2^{ND}|_{TE-TM}$ is plotted against the sum of two-photon energies normalized to $E_g + E_{hh}^{11}$.

From Fig. 10 we observe that ND-2PA in the TE-TM case is also structured as in the TM-TM case. Here the 1$^{st}$ transition starts when the sum of the two-photon energies becomes greater than $E_g + E_{hh}^{11}$. For $(\hbar\omega_1 + \hbar\omega_2)/E_g = 1.02$ in the bulk case, $\alpha_2^{ND}(\omega_1; \omega_2) \approx 82$ cm/GW and for $(\hbar\omega_1 + \hbar\omega_2)/E_g + E_{lh}^{11} = 1.02$ in the QW of width 10 nm, $\alpha_2^{ND}|_{TE-TM}(\omega_1; \omega_2) \approx 110$ cm/GW. Since the two-photon transition energy starts at $E_g + E_{hh}^{11}$ the ratio of $(\hbar\omega_1 + \hbar\omega_2)$ and $E_g + E_{lh}^{11}$ is taken to compare the ND-2PA coefficients of the bulk and QW for the TE-TM case on the same scale. As we see, the values of ND-2PA in the QW for the TE-TM case and the bulk case are almost equal. As we go deeper into the band, the ND-2PA in the bulk case is even larger than in the TE-TM case for the QW.

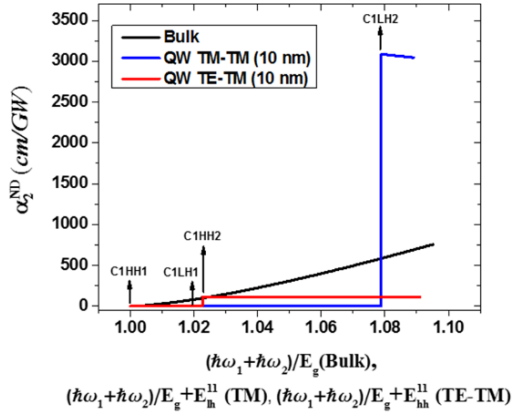

Fig. 10. ND-2PA of bulk GaAs and GaAs QW of width 10 nm for TM-TM and TE-TM cases for pump photon energy $\hbar\omega_2$ (7.5 μm).

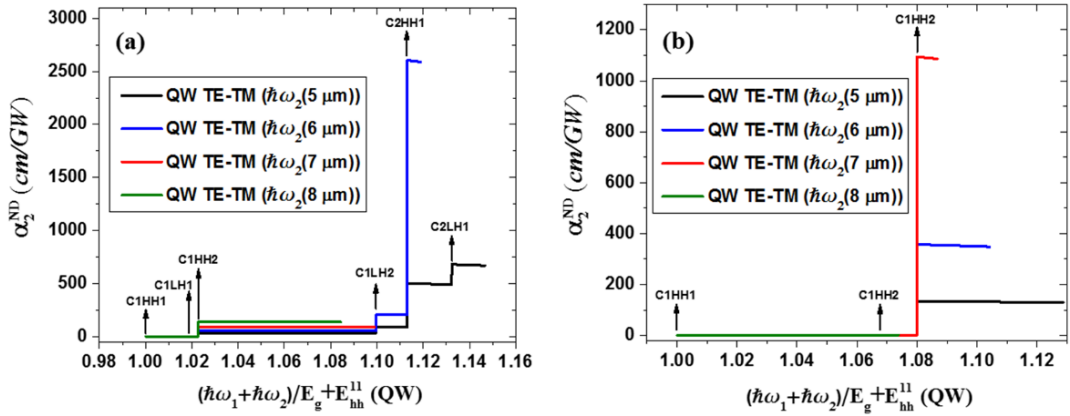

Fig. 11. (a) ND-2PA coefficient ND-2PA of GaAs QW of width (a) 10 $nm$ (b) 5 $nm$ for the TE-TM case. The arrows indicate valence band to conduction band transition energies.

For a better understanding of ND-2PA in the TE-TM case, $\alpha_2^{ND}|_{TE-TM}(\omega_1; \omega_2)$ for QW's of different widths is plotted in Fig. 11 for several pump photon energies. For a QW of width 10 $nm$ (Fig. 11 (a)) we observe that $\alpha_2^{ND}|_{TE-TM}(\omega_1; \omega_2) \approx 2600$ $cm/GW$ at the C2HH1 transition for a pump photon energy of $\hbar\omega_2$ (6 $\mu m$). For other pump photon energies $\hbar\omega_2$ (5 μm), $\hbar\omega_2$ (7 μm), and $\hbar\omega_2$ (8 μm) we did not observe a strong enhancement in $\alpha_2^{ND}|_{TE-TM}(\omega_1; \omega_2)$.

For a QW of width 5 $nm$ (Fig. 11 (b)), for pump photon energies $\hbar\omega_2$ (5 $\mu m$), $\hbar\omega_2$ (6 $\mu m$), and $\hbar\omega_2$ (7 $\mu m$), we observe that $\alpha_2^{ND}|_{TE-TM}(\omega_1;\omega_2)$ is not highly enhanced in comparison to the bulk. For the pump photon energy, $\hbar\omega_2$ (8 $\mu m$) 2PA is not energetically allowed because $\hbar\omega_2 < E_{hh}^2 - E_{hh}^1$. In the TE-TM case the probe photon energy is limited by the relation $\hbar\omega_1 < E_g + E_{hh}^{11}$ resulting in a narrow range of photon energies to obtain enhancement of ND-2PA over the bulk.

## V.  ENHANCEMENT OF TWO-PHOTON CARRIER GENERATION IN TWO-PHOTON DETECTION

A possible application of the enhancement of 2PA in the extreme nondegenerate case is IR detection using uncooled wide-bandgap photodiodes [10]. For pulsed IR detection, a gate pulse of photon energy slightly less than the bandgap energy is used to sensitize the photodiode, during which the IR pulse is detected. This has also been demonstrated to work for detection of continuous-wave radiation [11]. Fishman et al, [10] demonstrated gated detection of sub−100 $pJ$ mid-infrared radiation using an uncooled GaN detector. For gated IR detection using ND-2PA, the signal measured at the output of the photodetector is proportional to the photo-generated carrier density $N$, which is given by

$$\frac{dN}{dt} = \frac{dN_{ND}}{dt} + \frac{dN_D}{dt} = 2\frac{\alpha_2^{ND}(\omega_1;\omega_2)}{\hbar\omega_1}I_2 I_1 + \frac{\alpha_2^D(\omega_2;\omega_2)}{2\hbar\omega_2}I_2^2, \quad (56)$$

$dN_{ND}/dt$ and $dN_D/dt$ are the carrier generation rates for the ND-2PA of the gate and IR photons and the D-2PA of the gate photons respectively. The photo-generated carrier density due to the ND-2PA term can be written as

$$\frac{dN_{ND}}{dt} = 2\frac{\alpha_2^{ND}(\omega_1;\omega_2)}{\hbar\omega_1}I_2 I_1 = 2\frac{\alpha_2^{ND}(\omega_2;\omega_1)}{\hbar\omega_2}I_1 I_2. \quad (57)$$

We observe that the signal at the output of the detector, or the carrier generation rate, is proportional to $\alpha_2^{ND}(\omega_1;\omega_2)/\hbar\omega_1 = \alpha_2^{ND}(\omega_2;\omega_1)/\hbar\omega_2$.

For IR detection purposes it is instructive to compare the carrier generation enhancement of ND-2PA in QW's with ND-2PA in bulk semiconductors. Fig. 12 (a) shows a plot of $\alpha_2^{ND}(\omega_1;\omega_2)/\hbar\omega_1$ as a function of the normalized photon energies $\hbar\omega_1/E_g$ and $\hbar\omega_2/E_g$ for the bulk, and Fig. 12 (b) shows a plot of $\alpha_2^{ND}|_\perp(\omega_1;\omega_2)/\hbar\omega_1$ of a QW of width 10 $nm$, as a function of the normalized photon energies $\hbar\omega_1/(E_g + E_{lh}^{11})$ and $\hbar\omega_2/(E_g + E_{lh}^{11})$ for the TM-TM case. The shaded triangular area corresponds to regions where 2PA occurs. The dotted line corresponds to the degenerate case and the arrows show the extreme nondegenerate case. These plots enable us to determine the best experimental conditions for detection using END photon pairs in bulk and QW semiconductors.

To compare the carrier generation rate we plot (Fig. 13) the $\alpha_2^{ND}(\omega_1;\omega_2)/\hbar\omega_1$ values from Fig. 12 as a function of the nondegeneracy ($\hbar\omega_1/\hbar\omega_2$), for the bulk and the QW, where the $\alpha_2^{ND}(\omega_1;\omega_2)/\hbar\omega_1$ for the bulk is taken at the sum of the two-photon energies  $\hbar\omega_1 + \hbar\omega_2 = 1.112 E_g$ and for the TM-TM case in the QW $\hbar\omega_1 + \hbar\omega_2 = 1.112(E_g + E_{lh}^{11})$.

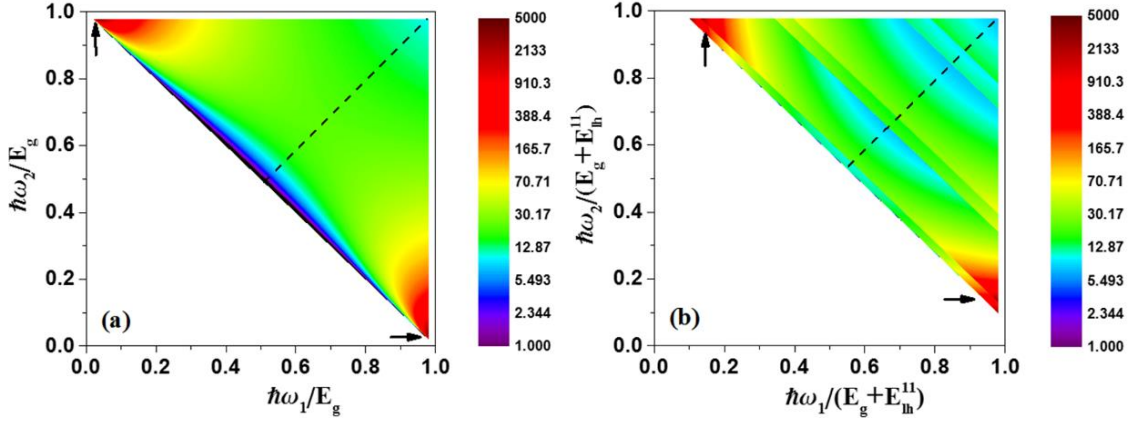

Fig. 12. (a) Theoretical predicted carrier generation enhancement for ND-2PA: (a) bulk semiconductors, (b) QW semiconductors of width 10 $nm$ for TM-TM. The arrows point to regions of large enhancement. The color scale is logarithmic.

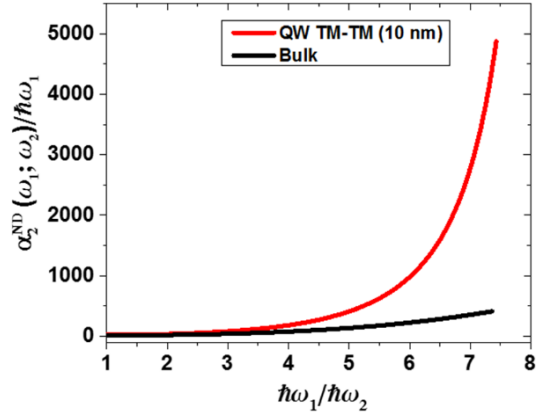

Fig. 13. Theoretical prediction of $\alpha_2^{ND}(\omega_1;\omega_2)/\hbar\omega_1$ for bulk GaAs and a GaAs QW of width 10 $nm$ in the TM-TM case.

Fig. 13 shows that the maximum enhancement of the carrier generation rate $\alpha_2^{ND}(\omega_1;\omega_2)/\hbar\omega_1$ observed in the TM-TM case of the QW is $\approx 12$ times that in bulk. In principle, this enhancement can be made larger by moving closer to $E_g + E_{lh}^{11}$, but we run into linear absorption losses. Here we restricted $\hbar\omega_1$ to 30 meV below the linear absorption edge. Urbach tail absorption of the "pump" radiation becomes an important factor as the pump photon energy approaches the bandgap. In the photo-detector applications this absorption will cause "effective" dark current. Therefore, for practical application the ratio of pump/probe photons energies should remain below ~1:10 to avoid absorption of the pump.

## VI. CONCLUSION

In this paper we used perturbation theory to calculate degenerate two-photon absorption (D-2PA) and nondegenerate two-photon absorption (ND-2PA) spectra in semiconductor quantum wells. Analytical expressions are obtained for the 2PA spectra for all possible permutations of polarizations. The most significant resonant enhancement of ND-2PA is possible for TM-TM polarized light. The results showed an order of magnitude enhancement in the ND-2PA coefficient in the TM-TM case of the QW over the bulk case.

In ND-2PA for bulk semiconductors the enhancement occurs due to the energies of the individual photons involved approaching the interband and intraband resonances. Thus the enhancement is maximized when $\hbar\omega_1$ nears $E_g$ and $\hbar\omega_2$ approaches zero. However, in this limit the density of states also approaches zero and the intraband matrix element, which varies linearly with $k$ as described by (19), also approaches zero, and the intermediate state resonance enhancement is thus limited.

In QW's, for the TM-TM case, the enhancement is due to interband and intersubband resonances. These occur when $\hbar\omega_1$ approaches $E_g + E_{lh}^{11}$ (interband) and when $\hbar\omega_2$ approaches $E_c^2 - E_c^1$ or $\hbar\omega_2$ goes to $E_{lh}^2 - E_{lh}^1$ (intersubband). As we approach these resonances both the density of states and the transition matrix elements remain finite. In QW's the larger density of states near the band edge also enhances the ND-2PA over that for the bulk semiconductor.

The calculations presented here are useful for determining the potential of QW's for nonlinear optical devices. The large enhancement in ND-2PA obtained for QW's should enable many new applications. The strong enhancement for the TM-TM polarization can be used for mid-IR detection using END-2PA with QW's acting as the active region in a photodiode. Apart from applications in detection, the enhancement can be used for all-optical switching using microring resonators and waveguides with direct-bandgap semiconductor QW layers as the core. For microring resonators the resonating light is just below the linear absorption edge and the cavity Q can be easily spoiled via END-2PA by using pump pulses in the mid-IR. As two-photon gain is the inverse of 2PA, using inverted QW's may provide two-photon gain and a possible path to a semiconductor two-photon laser [10], [48], [56], [70]–[73].

### ACKNOWLEDGEMENTS

This work was supported by the National Science Foundation grants ECCS-1202471 and ECCS-1229563.

The authors Himansu S. Pattanaik, David J. Hagan, and Eric W. VanStryland are with the CREOL, The College of Optics and Photonics, University of Central Florida, Orlando, FL 32816 USA (e-mail: himansu@creol.ucf.edu; hagan@creol.ucf.edu; ewvs@creol.ucf.edu).
Matthew Reichert, was with the CREOL, The College of Optics and Photonics, University of Central Florida, Orlando, FL 32816 USA. He is now with the Department of Electrical Engineering, Princeton University, Princeton, NJ 08455 USA (e-mail: mr22@princeton.edu).
Jacob B. Khurgin is with the Department of Electrical and Computer Engineering, The Johns Hopkins University, Baltimore, MD 21218 USA (e-mail: jakek@jhu.edu).

### REFERENCES

[1] A. Rostami, "Low threshold and tunable all-optical switch using two-photon absorption in array of nonlinear ring resonators coupled to MZI," *Microelectronics J.*, vol. 37, pp. 976–981, 2006.
[2] S. Hendrickson, C. Weiler, R. Camacho, P. Rakich, a. Young, M. Shaw, T. Pittman, J. Franson, and B. Jacobs, "All-optical-switching demonstration using two-photon absorption and the Zeno effect," *Phys. Rev. A*, vol. 87, p. 023808, 2013.
[3] I. S. Maksymov, L. F. Marsal, and J. Pallarès, "Modeling of two-photon absorption in nonlinear photonic crystal all-optical switch," *Opt. Commun.*, vol. 269, pp. 137–141, 2007.
[4] B. Jacobs and J. Franson, "All-optical switching using the quantum Zeno effect and two-photon absorption," *Phys. Rev. A*, vol. 79, no. 6, pp. 1–10, Jun. 2009.
[5] H. K. Tsang, R. S. Grant, R. V. Penty, I. H. White, J. B. D. Soole, H. P. LeBlanc, N. C. Andreadakis, M. S. Kim, and W. Sibbett, "GaAs/GaAlAs multiquantum well waveguides for all-optical switching at 1.55μm," *Electron. Lett.*, vol. 27, no. 22, pp. 1993–1995, 1991.
[6] E. W. Van Stryland, Y. Y. Wu, D. J. Hagan, M. J. Soileau, and K. Mansour, "Optical limiting with semiconductors," *J. Opt. Soc. Am. B*, vol. 5, no. 9, p. 1980, 1988.
[7] E. W. Van Stryland, H. Vanherzeele, M. A. Woodall, M. J. Soileau, A. L. Smirl, S. Guha, and T. F. Boggess, "Two Photon Absorption, Nonlinear Refraction, And Optical Limiting In Semiconductors," *Opt. Eng.*, vol. 24, no. 4, pp. 613–623, 1985.
[8] D. Duchesne, L. Razzari, L. Halloran, R. Morandotti, A. J. SpringThorpe, D. N. Christodoulides, and D. J. Moss, "Two-photon detection in a MQW GaAs laser at 1.55μm," *2009 Conf. Lasers Electro-Optics 2009 Conf. Quantum Electron. Laser Sci. Conf.*, no. Figure 1, pp. 5–6, 2009.
[9] D. Duchesne, L. Razzari, L. Halloran, M. Giguere, F. Legare, R. Morandotti, a. J. SpringThorpe, D. N. Christodoulides, and D. J. Moss, "Two-photon Autocorrelation in a MQW GaAs Laser at 1.55μm," *PIERS Online*, vol. 6, no. 1, pp. 267–272, 2010.

[10] D. A. Fishman, C. M. Cirloganu, S. Webster, L. A. Padilha, M. Monroe, D. J. Hagan, and E. W. Van Stryland, "Sensitive mid-infrared detection in wide-bandgap semiconductors using extreme non-degenerate two-photon absorption," *Nat. Photonics*, vol. 5, no. 9, pp. 561–565, Aug. 2011.

[11] H. S. Pattanaik, D. A. Fishman, D. J. Hagan, and E. W. Van Stryland, "Pulsed and CW IR Detection in Wide-gap Semiconductors using Extremely Nondegenerate Two-photon Absorption," in *Conference on Lasers and Electro-Optics/International Quantum Electronics Conference*, 2013, no. 1, pp. 3–4.

[12] M. Göppert-Mayer, "Uber Elementarakte mit zwei Quantensprüngen," *Ann. Phys.*, vol. 401, pp. 273–294, 1931.

[13] R. Cingolani, M. Lepore, R. Tommasi, I. M. Catalano, H. Lage, U. Gneqe, U. Bari, and K. Amendola, "Two-photon absorption in low-dimensional heterostructures," *J. Phys. IV*, vol. 3, pp. 131–138, 2000.

[14] I. M. Catalano, M. Lepore, R. Cingolani, and K. Ploog, "Two-Photon Absorption Processes in GaAs / Alx Ga1- xAs Quantum Wells (*).," *NUOVO Cim.*, vol. 12 D, no. 10, 1990.

[15] E. CUMBERBATCH, "Self-focusing in Non-linear Optics," *IMA J. Appl. Math.*, vol. 6, no. 3, pp. 250–262, 1970.

[16] R. Y. Chiao, T. K. Gustafson, and P. L. Kelley, "Self-focusing of optical beams," *Top. Appl. Phys.*, vol. 114, no. 26, pp. 129–143, 2009.

[17] F. Shimizu, "Frequency broadening in liquids by a short light pulse," *Phys. Rev. Lett.*, vol. 19, no. 19, pp. 1097–1100, 1967.

[18] R. Stolen and C. Lin, "Self-phase-modulation in silica optical fibers," *Phys. Rev. A*, vol. 17, no. 4, pp. 1448–1453, 1978.

[19] M. N. Islam, J. R. Simpson, H. T. Shang, L. F. Mollenauer, and R. H. Stolen, "Cross-phase modulation in optical fibers," *Opt. Lett.*, vol. 12, no. 8, pp. 625–627, 1987.

[20] P. S. Spencer and K. a. Shore, "Pump—probe propagation in a passive Kerr nonlinear optical medium," *J. Opt. Soc. Am. B*, vol. 12, no. 1, p. 67, 1995.

[21] R. Stolen, "Phase-matched-stimulated four-photon mixing in silica-fiber waveguides," *IEEE J. Quantum Electron.*, vol. 11, pp. 100–103, 1975.

[22] R. Stolen and J. Bjorkholm, "Parametric amplification and frequency conversion in optical fibers," *IEEE J. Quantum Electron.*, vol. 18, no. 7, pp. 1062–1072, 1982.

[23] T. Tekin, M. Schlak, W. Brinker, B. Maul, and R. Molt, "Monolithically integrated MZI comprising band gap shifted SOAs: a new switching scheme for generic all-optical signal processing," *Eur. Conf. Opt. Commun.*, no. August, pp. 123–124, 2000.

[24] T. Ohara, H. Takara, I. Shake, K. Mori, S. Kawanishi, S. Mino, T. Yamada, M. Ishii, T. Kitoh, T. Kitagawa, K. R. Parameswaran, and M. M. Fejer, "160-Gb/s optical-time-division multiplexing with PPLN hybrid integrated planar lightwave circuit," *IEEE Photonics Technol. Lett.*, vol. 15, no. 2, pp. 302–304, 2003.

[25] X. Yang, A. K. Mishra, D. Lenstra, F. M. Huijskens, H. De Waardt, G. D. Khoe, and H. J. S. Dorren, "Sub-picosecond all-optical switch using a multi-quantum-well semiconductor optical amplifier," *Opt. Commun.*, vol. 236, pp. 329–334, 2004.

[26] M. Sheik-bahae, D. C. Hutchings, D. J. Hagan, E. W. Van Stryland, and S. Member, "Dispersion of Bound Electronic Nonlinear Refraction in Solids," *IEEE J. Quantum Electron.*, vol. 27, no. 6, pp. 1296–1309, 1991.

[27] B. S. Wherrett, "Scaling rules for multiphoton interband absorption in semiconductors," *J. Opt. Soc. Am. B*, vol. 1, no. 1, p. 67, Mar. 1984.

[28] E. W. Van Stryland, M. A. Woodall, H. Vanherzeele, and M. J. Soileau, "Energy band-gap dependence of two-photon absorption," *Opt. Lett.*, vol. 10, no. 10, p. 490, Oct. 1985.

[29] H. S. Brandi and C. B. De Araujos, "Multiphonon absorption coefficients in solids: a universal curve," *J. Phys. C Solid State Phys.*, vol. 16, pp. 5929–5936, 2000.

[30] V. E. W. Hutchings, D C, "Nondegenerate two-photon absorption in zinc blende semiconductors," *J. Opt. Soc. Am. B*, vol. 9, no. 11, 1992.

[31] Y. Arakawa and A. Yariv, "Quantum Well Lasers--Gain, Spectral, Dynamics," *IEEE J. Quantum Electron.*, vol. QE-22, no. 9, pp. 1887–1899, 1986.

[32] S. Kalchmair, R. Gansch, S. I. Ahn, A. M. Andrews, H. Detz, T. Zederbauer, E. Mujagić, P. Reininger, G. Lasser, W. Schrenk, and G. Strasser, "Detectivity enhancement in quantum well infrared photodetectors utilizing a photonic crystal slab resonator.," *Opt. Express*, vol. 20, no. 5, pp. 5622–8, Feb. 2012.

[33] T. H. Wood, C. A. Burrus, D. A. B. Miller, D. S. Chemla, T. C. Damen, A. C. Gossard, and W. Wiegmann, "High-speed optical modulation with GaAs/GaAlAs quantum wells in a p-i-n diode structure," *Appl. Phys. Lett.*, vol. 44, no. 1, p. 16, 1984.

[34] D. Ahn, L. C. Kimerling, and J. Michel, "Efficient evanescent wave coupling conditions for waveguide-integrated thin-film Si∕Ge photodetectors on silicon-on-insulator∕germanium-on-insulator substrates," *J. Appl. Phys.*, vol. 110, no. 8, p. 083115, 2011.

[35] S. Hinds, M. Buchanan, R. Dudek, S. Haffouz, S. Laframboise, Z. Wasilewski, and H. C. Liu, "Near-room-temperature mid-infrared quantum well photodetector.," *Adv. Mater.*, vol. 23, no. 46, pp. 5536–9, Dec. 2011.

[36] A. G. U. Perera, W. Z. Shen, S. G. Matsik, H. C. Liu, M. Buchanan, and W. J. Schaff, "GaAs/AlGaAs quantum well photodetectors with a cutoff wavelength at 28μm," *Appl. Phys. Lett.*, vol. 72, no. 13, pp. 1596–1598, 1998.

[37] T. H. Wood, "Multiple quantum well waveguide modulators," *J. Light. Technol*, vol. vol, no. 6, pp. 6pp743–757, 1988.

[38] G. D. Boyd, D. A. B. Miller, and D. S. Chamela, "Multiple quantum well reflection modulator.pdf," *Appl. Phys. Lett.*, vol. 50, no. 17, pp. 1119–1121, 1987.

[39] D. Duchesne, L. Razzari, L. Halloran, R. Morandotti, A. J. SpringThorpe, D. N. Christodoulides, and D. J. Moss, "Two-photon photodetector in a multiquantum well GaAs laser structure at 1.55μm," *Opt. Express*, vol. 17, no. 7, pp. 5298–5310, 2009.

[40] J. U. Kang, J. B. Khurgin, C. C. Yang, H. H. Lin, and G. I. Stegeman, "Two-photon transitions between bound-to-continuum states in AlGaAs/GaAs multiple quantum well," *Appl. Phys. Lett.*, vol. 73, no. 1998, pp. 3638–3640, 1998.

[41] S. Li and J. B. Khurgin, "Two photon confined-to-continuum intersubband transitions in the semiconductor heterostructures," *J. Appl. Phys.*, vol. 73, no. 1993, pp. 4367–4369, 1993.

[42] J. B. Khurgin, "Nonlinear response of the semiconductor quantum-confined structures near and below the middle of the band gap," *J. Opt. Soc. Am. B*, vol. 11, no. 4, p. 624, Apr. 1994.

[43] C. . C. Yang, A. Villeneuve, G. I. Stegeman, C. Lin, and H. Lin, "Anisotropic two-photon transitions in GaAs/AlGaAs multiple quantum well waveguides," *IEEE J. Quantum Electron.*, vol. 29, no. 12, pp. 2934–2939, 1993.

[44] M. N. Islam, C. E. Soccolich, R. E. Slusher, a. F. J. Levi, W. S. Hobson, and M. G. Young, "Nonlinear spectroscopy near half-gap in bulk and quantum well GaAs/AlGaAs waveguides," *J. Appl. Phys.*, vol. 71, no. 1992, pp. 1927–1935, 1992.

[45] S. Wagner, J. Meier, A. Helmy, J. Aitchison, M. Sorel, and D. Hutchings, "Polarization-dependent nonlinear refraction and two-photon absorption in GaAs/AlAs superlattice waveguides below the half-bandgap," *J. Opt. Soc. Am. B*, vol. 24, no. 7, pp. 1557–1563, 2007.

[46] B. Portier, B. Vest, F. Pardo, N. Péré-Laperne, E. Steveler, J. Jaeck, C. Dupuis, N. Bardou, A. Lemaître, E. Rosencher, R. Haïdar, and J.-L. Pelouard, "Resonant metallic nanostructure for enhanced two-photon absorption in a thin GaAs p-i-n diode," *Appl. Phys. Lett.*, vol. 105, no. 2007, p. 011108, 2014.

[47] F. Boitier, A. Godard, J. Bonnet, E. Rosencher, and C. Fabre, "Measuring photon bunching at ultrashort timescale by two-photon absorption in semiconductors," *Nat. Phys.*, vol. 5, no. 4, pp. 267–270, 2009.

[48] A. Hayat, P. Ginzburg, and M. Orenstein, "Observation of two-photon emission from semiconductors," *Nat. Photonics*, vol. 2, no. April, pp. 238–241, 2008.

[49] A. Hayat, P. Ginzburg, and M. Orenstein, "High-rate entanglement source via two-photon emission from semiconductor quantum wells," *Phys. Rev. B - Condens. Matter Mater. Phys.*, vol. 76, no. 3, pp. 1–15, 2007.

[50] H. Spector, "Two-photon absorption in semiconducting quantum-well structures.," *Phys. Rev. B. Condens. Matter*, vol. 35, no. 11, pp. 5876–5879, Apr. 1987.

[51] A. Pasquarello and A. Quattropani, "Gauge-invariant two-photon transitions in quantum wells.," *Phys. Rev. B. Condens. Matter*, vol. 38, no. 9, pp. 6206–6210, Sep. 1988.

[52] K. Tai, A. Mysyrowicz, R. Fischer, R. Slusher, and A. Cho, "Two-photon absorption spectroscopy in GaAs quantum wells.," *Phys. Rev. Lett.*, vol. 62, no. 15, pp. 1784–1787, Apr. 1989.

[53] H. K. Tsang, R. V. Penty, I. H. White, R. S. Grant, W. Sibbett, J. B. D. Soole, H. P. LeBlanc, N. C. Andreadakis, E. Colas, and M. S. Kim, "Polarization and field dependent two-photon absorption in GaAs/AlGaAs multiquantum well waveguides in the half-band gap spectral region," *Appl. Phys. Lett.*, vol. 59, no. 26, p. 3440, 1991.

[54] C. Xia and H. N. Spector, "Nonlinear Franz – Keldysh effect : two-photon absorption in a semiconducting quantum well," *J. Opt. Soc. Am. B*, vol. 27, no. 8, pp. 1571–1575, 2010.

[55] A. Obeidat and J. Khurgin, "Excitonic enhancement of two-photon absorption in semiconductor quantum-well structures," *J. Opt. Soc. Am. B*, vol. 12, no. 7, p. 1222, 1995.

[56] C. M. Cirloganu, L. A. Padilha, D. A. Fishman, S. Webster, D. J. Hagan, and E. W. Van Stryland, "Extremely nondegenerate two-photon absorption in direct-gap semiconductors [Invited].," *Opt. Express*, vol. 19, no. 23, pp. 22951–60, Nov. 2011.

[57] H. S. Pattanaik, D. A. Fishman, S. Webster, D. J. Hagan, and E. Van Stryland, "IR detection in wide-gap semiconductors using extreme nondegenerate two-photon absorption," *Conf. Lasers Electro-Optics 2012*, vol. 2, p. QF2G.7, 2012.

[58] J. M. Hales, D. J. Hagan, E. W. Van Stryland, K. J. Schafer, A. R. Morales, K. D. Belfield, P. Pacher, O. Kwon, E. Zojer, and J. L. Bredas, "Resonant enhancement of two-photon absorption in substituted fluorene molecules," *J. Chem. Phys.*, vol. 121, no. 2004, pp. 3152–3160, 2004.

[59] S. H. Lin, *Multiphoton Spectroscopy of Molecules*. 1984.

[60] G. J. Brown and F. Szmulowicz, "Normal Incidence Detection of Infrared radiation in P-type GaAs/AlGaAs Quantum Well Structures," in *Long Wavelength Infrared detectors*, Amsterdam: Manijeh Razeghi, 1996, pp. 271–333.

[61] J. M. Luttinger, "Quantum Theory of Cyclotron Resonance in Semiconductors-General Theory," *Phys. Rev.*, vol. 102, no. 4, pp. 1030–1041, 1956.

[62] E. O. Kane, "Band Structure of Indium Antimonide," *J. Phys. Chem. Solids*, vol. 1, pp. 249–261, 1957.

[63] D. C. Hutchings and E. W. Van Stryland, "Nondegenerate two-photon absorption in zinc blende semiconductors," *J. Opt. Soc. Am. B*, vol. 9, no. 11, pp. 2065–2074, 1992.

[64] H. D. Jones and H. R. Reiss, "Intense-field effects in solids," *Phys. Rev. B*, vol. 16, no. 6, pp. 2466–2473, 1977.

[65] L. V. Keldysh, "Ionization in the field of a string electromagnetic wave," *J. Exp. Theor. Phys.*, vol. 20, no. 5, pp. 1307–1314, 1965.

[66] A. Shimizu, "Two-photon absorption in quantum-well structures near half the direct band gap.," *Phys. Rev. B. Condens. Matter*, vol. 40, no. 2, pp. 1403–1406, Jul. 1989.

[67] H. S. Pattanaik, M. Reichert, H. Hu, D. J. Hagan, and E. W. Van Stryland, "Scanning 3-D IR Imaging With a GaN Photodiode Using Nondegenerate Two-photon Absorption," in *Frontiers in Optics 2013 Postdeadline*, 2013, p. FW6C.7.

[68] H. Pattanaik, D. Fishman, S. Webster, D. Hagan, and E. Van Stryland, "IR detection in wide-gap semiconductors using extreme nondegenerate two-photon absorption," in *Conference on Lasers and Electro-Optics 2012*, 2012, p. QF2G.7.

[69] H. S. Pattanaik, D. A. Fishman, D. J. Hagan, and E. W. Van Stryland, "CW IR Detection in Wide-gap Semiconductors Using Extremely Nondegenerate Two-photon Absorption," in *Frontiers in Optics 2013/Laser Science XXIX*, 2013, p. FTh4C.3.

[70] C. N. Ironside, "Two-photon gain semiconductor amplifier," *IEEE J. Quantum Electron.*, vol. 28, no. 4, pp. 842–847, 1992.

[71] D. J. Gauthier, Q. Wu, S. E. Morin, and T. W. Mossberg, "Realization of a continuous-wave, two-photon optical laser," *Phys. Rev. B - Condens. Matter Mater. Phys.*, vol. 68, no. 4, pp. 464–467, 1992.

[72] A. Nevet, A. Hayat, and M. Orenstein, "Measurement of Optical Two-Photon Gain in Electrically Pumped AlGaAs at Room Temperature," *Phys. Rev. Lett.*, vol. 104, no. 20, p. 207404, 2010.

[73] D. J. Gauthier, "Two-photon lasers," *Prog. Opt.*, vol. 45, pp. 205–268, 2003.

[74] S. L. Chuang, *Physics of Optoelectronic Devices*. 1995.

[75] S. W. Corzine, R.-H. Yan, and L. A. Coldren, *Quantum Well Lasers*. 1993.

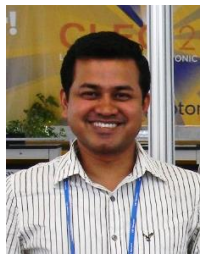
**Himansu S. Pattanaik** received his M. Tech degree in Applied Optics at Indian Institute of Technology, Delhi and PhD Degree in Optics and Photonics at CREOL, The College of Optics & Photonics, University of Central Florida, FL, USA in 2015.

He is currently working as a Post-Doctoral Associate in the Nonlinear Optics Group at CREOL, The College of Optics & Photonics, University of Central Florida. His research interests include nonlinear optics and its application to novel photonic devices.

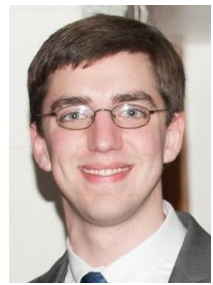
**Matthew Reichert** received his B.S. degree in Optical Engineering from Rose-Hulman Institute of Technology in 2010, and Ph.D. in Optics and Photonics from CREOL, The College of Optics and Photonics, University of Central Florida in 2015.

He is currently a Postdoctoral Research Associate in the Department of Electrical Engineering at Princeton University. His research interests include nonlinear optical spectroscopy and quantum imaging.

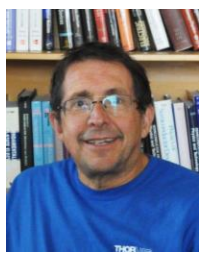
**Jacob B. Khurgin** received his PhD from Polytechnic Institute of New York University in 1987. Prior to that he worked at Philips NV Laboratories in Briarcliff Manor, NY where his interest included solid state lasers, semiconductor lasers, nonlinear optics, displays, and lighting. Since 1988 he has been a Professor of Electrical and Computer Engineering at Johns Hopkins University in Baltimore MD where he worked in the fields of opto-electronics, condensed matter physics, optical communications, microwave photonics, nonlinear optics, slow light, plasmonics, and so on.

He as an author of about 300 refereed papers, 6 book chapters and 34 US patents. Prof Khurgin is an APS and OSA Fellow.

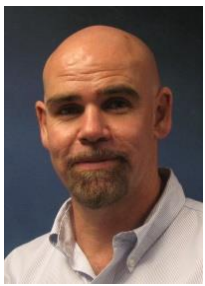
**David J. Hagan** received his PhD degree in Physics at Heriot-Watt University, Edinburgh, Scotland in 1985. He is a Professor of Optics and Photonics and Associate Dean for Academic Programs of the College of Optics and Photonics, and is an OSA Fellow. He is also the founding Editor-in-Chief of the journal, Optical Materials Express.

His research interests include techniques for nonlinear optical materials characterization, optical power limiting and switching, and methods for enhancement of optical nonlinearities.

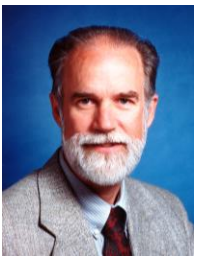
**Eric W. Van Stryland** received his physics PhD working at the Optical Sciences Center, University of Arizona, 1976 and joined the University of North Texas. He joined the start of CREOL-1987, become director-1999, and its first Dean-2004, The College of Optics and Photonics.

Dr. VanStryland is past President of OSA and Fellow of OSA (R.W. Wood Prize winner), SPIE, IEEE, APS and past Board member of LIA. He graduated 37 Ph.D.'s, published >300 papers primarily in the field of nonlinear optics (e.g. Z-scan, nonlinear Kramers-Kronnig, cascaded second-order nonlinearities), and is Pegasus Professor and Trustee Chair.